\documentclass[aps,prc,onecolumn,groupedaddress]{revtex4-2}

\usepackage{graphicx}

\usepackage{bm}
\usepackage{amsmath}
\usepackage{aas_macros}
\usepackage{xcolor}

\graphicspath{{./}{figs/}}

\begin{document}

\title{Bayesian averaging for ground state masses of atomic nuclei in a Machine Learning approach}

\author{M.~R.~Mumpower}
\email[]{mumpower@lanl.gov}
\affiliation{Theoretical Division, Los Alamos National Laboratory, Los Alamos, NM, 87545, USA}

\author{M.~Li}
\affiliation{Clemson University, Department of Physics and Astronomy, Clemson, SC 29634-0978, USA}
\affiliation{Theoretical Division, Los Alamos National Laboratory, Los Alamos, NM, 87545, USA}

\author{T.~M.~Sprouse}
\affiliation{Theoretical Division, Los Alamos National Laboratory, Los Alamos, NM, 87545, USA}

\author{B.~S.~Meyer}
\affiliation{Clemson University, Department of Physics and Astronomy, Clemson, SC 29634-0978, USA}

\author{A.~E.~Lovell}
\affiliation{Theoretical Division, Los Alamos National Laboratory, Los Alamos, NM, 87545, USA}

\author{A.~T.~Mohan}
\affiliation{Theoretical Division, Los Alamos National Laboratory, Los Alamos, NM, 87545, USA}

\date{\today}

\begin{abstract}
We present global predictions of the ground state mass of atomic nuclei based on a novel Machine Learning (ML) algorithm. 
We combine precision nuclear experimental measurements together with theoretical predictions of unmeasured nuclei. 
This hybrid data set is used to train a probabilistic neural network. 
In addition to training on this data, a physics-based loss function is employed to help refine the solutions. 
The resultant Bayesian averaged predictions have excellent performance compared to the testing set and come with well-quantified uncertainties which are critical for contemporary scientific applications. 
We assess extrapolations of the model's predictions and estimate the growth of uncertainties in the region far from measurements. 
\end{abstract}

\maketitle

\section{Introduction}

Mass is a defining quantity of an atomic nucleus and appears ubiquitously in research efforts ranging from technical applications to scientific studies such as the synthesis of the heavy elements in astrophysical environments \citep{Horowitz2019, Kajino2019}. While accurate nuclear data of masses is available for nuclei that are relatively stable, the same is not true for nuclei farther away from beta stability because measurements on radioactive nuclei are exceedingly challenging \citep{Thoennessen2010}.   As a consequence, theoretical models of atomic nuclei are required for extrapolations used in present-day scientific applications \citep{Moller2003}. 

The goal of theoretical nuclear models is to describe all atomic nuclei (from light to heavy) using fundamental interactions.  Attainment of this challenging goal remains elusive, however, due to the sheer complexity of modeling many-body systems with Quantum Chromodynamics \citep{Geesaman2015}. 
To understand the range of nuclei that may exist in nature, mean-field approximations are often made which simplify complex many-body dynamics into a non-interacting system of quasi-particles where remaining residual interactions can be added as perturbations \citep{Negele1982}. 
A consequence of this approximation is that current nuclear modeling efforts are unable to describe the rich correlations that are found across the chart of nuclides.

In contrast, Machine Learning (ML) based approaches do not have to rely on the assumption of modeling nuclei from a mean-field. 
This provides freedom in finding solutions that contemporary modeling may not be capable of ascertaining. 
Furthermore, Bayesian approaches to ML afford the ability to associate predictions with uncertainties \citep{Goan2020, Mohebali2020}.
Such tasks are more difficult to achieve in modern nuclear modeling due to relatively higher computational costs. 

ML approaches in nuclear physics were pioneered by J.W.~Clark and colleagues \citep{Clark1991, Clark1992}. 
These studies were the first to show that networks could approximate stable nuclei, learn to predict masses and analyze nuclear systematics of separation energies as well as spin-parity assignments \citep{Gazula1992, Gernoth1993}. 
Powered by open-source frameworks, research into ML methods has seen a recent resurgence in nuclear physics \citep{Boehnlein2022}. 
ML approaches have shown promise in optimizing data and experiments \citep{Hutchinson2022}, building surrogate models of density functional theory \citep{Verriere2022}, and describing quantum many-body wave functions for light nuclei\citep{Adams2021, Gnech2022}. 

Several research groups are actively pursuing the problem of describing nuclear systems with ML from a more data-centric approach. 
These efforts currently attempt to improve existing nuclear models by adding correction terms \citep{Utama2016a}. 
Gaussian Processes (GP) have also been used for model averaging \citep{Neufcourt2019}, but this approach is inherently limited to where data is known as GP methods typically revert to the mean when extrapolating. 
A further limitation to training ML models on residuals (or the discrepancy of theoretical model predictions with experimental data) is that the methods are arbitrary. 
The changes learned by the network to improve one model will not be applicable to another. 
These approaches thus provide limited insight into the underlying missing physics in modern models of the atomic nucleus. 

In \citet{Lovell2022}, a different approach was taken, where the masses of atomic nuclei were modeled directly with a neural network. 
It was shown that the masses of nuclei can be well described, and model predictions with increased fidelity correlate strongly with a careful selection of physically motivated input features. 
The selection of input features is especially important in ML applications \citep{Niu2018, Perez2022}. 
Following this work, \citet{Mumpower2022} showed that the size of the training set can greatly be reduced, and the fidelity of model solutions increases drastically, when an additional physical constraint is introduced as a second loss function during model training.  

The focus of this work is to present a Bayesian approach for combining precision data with theoretical predictions to model the mass of atomic nuclei. 
In Section \ref{sec:methods}, we present our ML algorithm and define the model hyperparameters. 
In Section \ref{sec:results}, we show the results of our approach and assess the quality of model extrapolations. 
We end with a short summary. 

\section{Methods}
\label{sec:methods}

In this section, we outline our methodology: describe the neural network, define our physics-based feature space, list model hyperparameters, and discuss training. 

\subsection{Mixture Density Network}

In a feed-forward neural network, inputs, $\textbf{x}$, are mapped to outputs, $\textbf{y}$, in a deterministic manner. 
We employ the Mixture Density Network (MDN) of \citet{Bishop1994} which differs from the standard approach. 
This ML network takes as inputs stochastic realizations of probability distributions and maps this to a mixture of Gaussians. 
Thus, the network fundamentally respects the probabilistic nature of both known data and model predictions by both sampling the prior distribution of inputs and predicting the posterior distribution of the outputs. 

Formally, the conditional probability can be written as
\begin{equation}
    p(\textbf{y}|\textbf{x}) = \sum \limits _{i=1} ^K \pi _i (\textbf{x}) \mathcal{N} \left ( \textbf{y} | \mu _i (\textbf{x}), I \sigma _i (\textbf{x}) \right ) \ ,
\label{eqn:mdn_y}
\end{equation}
where $\mathcal{N}$ is the normal distribution with means, $\mu _i (\textbf{x})$, and standard deviations, $\sigma _i (\textbf{x})$. 
The $\pi_i (\textbf{x})$ represent the weighting of each Gaussian respectively. 
The covariance matrix is assumed to be diagonal, as indicated by the use of the notation $I \sigma _i$. 

The neural network outputs are $\bm{\pi}$, $\bm{\mu}$, and $\bm{\sigma}$ which depend only on the input training set information $\textbf{x}$ and the network weights. 
For ease of reading the equation we have kept the dependence of the network weights implicit.

The hyperbolic tangent function $a(z) = \frac{e^z-e^{-z}}{e^z+e^{-z}}$ is used as the activation function for the neurons in the linear layers of the network . 
At the final layer a softmax function is used for the $\pi_i$ so that the previous layer's output can be mapped to a vector that sums to unity. 
This choice ensures that the mixture of Gaussians can be safely interpreted as a probability. 
Our MDN uses the \texttt{PyTorch}~\citet{Paszke2019} framework and can be run on either CPU or GPU architectures. 

\subsection{Physics-based feature space}

We now discuss the components of the input vector, $\textbf{x}$. 
The ground state of an atomic nucleus comprises $Z$ protons and $N$ neutrons. 
While it is reasonable to start from these two independent features as inputs, \citep{Niu2018} and \citet{Lovell2022} reported that a modestly larger physics-based feature space drastically improves the prediction of masses. 
For this reason, we employ a combination of macroscopic and microscopic features that are of relevance to low-energy nuclear physics properties. 

In addition to the proton number $Z$ and neutron number $N$, we also use the mass number $A = Z + N$, and a measure of isospin asymmetry, $P_\textrm{asym} = \frac{N-Z}{A}$, as relevant macroscopic features. 
For the microscopic features that encode the quantized nature of atomic nuclei, we employ notions of pairing by considering the even-odd behavior of the proton, neutron, and mass numbers. 
This can calculated by observing the binary values of these quantities modulo 2; $Z_\textrm{eo} = Z \div 2$ , $N_\textrm{eo} = N \div 2$, $A_\textrm{eo} = A \div 2$. 
A notion of shell structure is also important. 
To encode this information we include the number of valence nucleons or holes (beyond the mid-shell) from the nearest major closed shell for protons, $V_p$, and neutrons, $V_n$, respectively. 
The value of $V_p$ or $V_n$ is zero at a closed shell and reaches a maximum at the mid-shell. 
The number of valence nucleons is correlated with more complex excitations in nuclei, including collective behavior that may appear \citep{Casten1987, Casten1999}. 
The closed proton shells are set to 8, 20, 28, 50, 82, and 114. 
The closed neutron shells are set to 8, 20, 28, 50, 82, 126 and 184. 
These choices are free parameters in our modeling and can be modified to explore different physics. 

The input feature space is then a nine component vector:
\begin{equation}
    \textbf{x} = (Z, N, A, P_\textrm{asym}, Z_\textrm{eo}, N_\textrm{eo}, A_\textrm{eo}, V_p, V_n) \ , 
\end{equation}
where the first four components can be considered macroscopic features and the last five are microscopic features. 
All remaining features beyond the second are functions of $Z$ and $N$ exclusively.  

\subsection{Hybrid data for training}

Our training is hybrid data consisting of two distinct input sets. 
The first is the mass data provided by the 2020 Atomic Mass Evaluation (AME2020) \citep{Wang2021}. 
The information in this set is very precise with an average reported mass uncertainty of roughly 25 keV. 
Modern experimental advances, such as Penning trap mass spectrometers enhanced with the Phase-Imaging Ion-Cyclotron-Resonance technique, enable such high precision measurements \citep{Nesterenko2018, Clark2023}. 

The second mass data are provided by modern theoretical models. 
The information in this set is less precise, owing to the approximations made in the modeling of atomic nuclei. 
This set can be calculated for nuclei that have not yet been measured, providing a valuable new source of information. 
Nuclear models in this second set include macroscopic-microscopic approaches like the Duflo-Zuker model \citep{Duflo1995}, the 2012 version of the Finite Range Droplet Model (FRDM) \citep{Moller2016}, the WS4 model \citep{Wang2014} and microscopic approaches like UNEDF \citep{Kortelainen2012}, and HFB32 \citep{Goriely2016}. 

In this work, we combine predictions from three theoretical models: FRDM2012, WS4, and HFB32. 
Because these models do not report individual uncertainties on their predictions, we instead estimate theoretical model uncertainty using the commonly quoted root-mean-square error or
\begin{equation}
    \sigma_\textrm{RMS} = \sqrt{ \frac{1}{N} \sum _j ^N (d_j - t_j)^2 } \ ,
\label{eqn:sigma_rms}
\end{equation}
where $d_j$ is the atomic mass from the AME and $t_j$ is the predicted atomic mass from the theoretical model. 
The sum runs over each $j$ which defines a nucleus, ($Z$,$N$). 
For FRDM2012, WS4, and HFB32, $\sigma_\textrm{RMS}$ = 0.606, 0.295 and 0.608 MeV respectively, using AME2020 masses. 
While the $\sigma_\textrm{RMS}$ is a good measure of overall model accuracy for measured nuclei, uncertainties are certainly larger for shorter-lived systems. 
In this work, we do not seek to preferentiate one model over another. 
For this reason, we increase the assumed uncertainty of WS4 to a more reasonable 0.500 MeV when probing its predicted masses further from stability. 

Training for the hybrid input data is taken at random, rather than selected based on any given criteria. 
The number of unique nuclei from experimental data is a free parameter in our training. 
The best performance is found for models provided with approximately 20\% of the AME, or 400 to 500 nuclei \citep{Mumpower2022}. 
The number of unique nuclei from theory is also a free parameter. 
We find that as few as 50 additional unmeasured nuclei can influence training, and therefore use this minimal number. 
In the case of theory data, we sample the masses of 50 randomly chosen nuclei from each of the three mass models independently. 

The benefit of using hybrid data is that the neural network is not limited to solutions of model averaging which can regress to the mean when extrapolating. 
Instead, the combination of hybrid data with ML-based methods affords the opportunity to create new models that are capable of reproducing data, capturing trends, and predicting yet to be measured masses with sound uncertainties. 

\subsection{Model training and hyperparameters}

The network is set up with 6 hidden layers and 10 hidden nodes per layer. 
The final layer turns the network into a Gaussian ad-mixture. 
For masses we choose a single Gaussian, although other physical quantities, such as fission yields, may require additional components \citep{Lovell2020}. 
The Adam optimizer is used with learning rate 0.0002 \citep{Kingma2017}. 
We also implement a weight-decay regularization with value 0.01. 
These hyperparameters were determined from a select set of runs where the values were varied. 

We perform model training with two loss functions. 
The first loss function, $\mathcal{L}_1$, captures the match to input data. 
The log-likelihood loss for data is written as,

\begin{equation}
\mathcal{L}_1 = - \ln {\left [ \sum \limits _{i=1} ^K \frac{\pi_i(\textbf{x})}{(2\pi)^{K/2} \sigma_i (\textbf{x})} \mathrm{exp} \left \{ -\frac{|| \textbf{y}-\mu_i(\textbf{x}) ||^2}{2\sigma_i (\textbf{x})^2} \right \} \right ]},
\label{eqn:ll_data}
\end{equation} 

where $\textbf{y}$ is the vector of training outputs and $K$ is the total number of Gaussian mixtures. 
The $\pi_i (\textbf{x})$, $\mu _i (\textbf{x})$, and $\sigma _i (\textbf{x})$, variables define the Gaussians, as in Eqn.~\ref{eqn:mdn_y}. 
The minimization of this loss function furnishes the posterior distributions of predicted masses. 

The hybrid mixture of experimental data and theoretical data enter into training as the variable \textbf{y}. 
Each nucleus defined uniquely by a proton number $Z$ and neutron number $N$.  
A Gaussian distribution is assumed to represent the probability distribution for sampling both experimental and theoretical data, 
\begin{equation}
\label{eqn:probe_mass}
f(y,\mu,\sigma) = \frac{1}{\sigma\sqrt{2\pi}} 
  \exp\left( -\frac{1}{2}\left(\frac{y-\mu}{\sigma}\right)^{\!2}\,\right) \ .
\end{equation}
For the high-precision experimental data taken from the AME, the mean is set to the evaluated mass, $\mu = M^\textrm{AME}_{Z,N}$, and the variance is set to the reported uncertainty of a nucleus' mass, $\sigma = \delta M^\textrm{AME}_{Z,N}$. 
For the theoretical data from the three mass models, the mean value is taken as the prediction of the given mass model respectively. 
The uncertainties in these models is not reported on a per nucleus basis. 
Therefore an approximation to the model's $\sigma_\textrm{RMS}$, which is computed with respect to the AME, is used as the variance in the probability distribution. 

In this work, we do not include masses of isomeric states in the training set. 
However, we note that since our previous works \citep{Lovell2022, Mumpower2022}, the AME2020 is now utilized, rather than the earlier AME2016. 
This data better refines the separation of ground state and isomeric states in evaluated masses, which continues to be a known source of systematic uncertainty in the evaluation of atomic masses.

For the AME data, we take roughly 500 nuclei for training, leaving the remaining 80\% of the AME as testing data. 
The number of stochastic realizations per nucleus is 50. 
For theory data, we take only 50 additional nuclei explicitly outside the AME. 
These nuclei are also taken at random.
A given nucleus is set to have 20 stochastic realizations per theory model, for a total of 60 samples overall.
From a set of testing runs, the above choices produce suitable models. 
We summarize the model hyperparameters in Table \ref{tab:hyperparams}. 
A more complete study of all model hyperparameters is the subject of future investigations. 

\begin{table}[]
\caption{The neural network hyperparameters used in this work. }
\label{tab:hyperparams}
\begin{tabular}{lll}
\textbf{Parameter}       & \textbf{Value(s)} & \textbf{Comment}  \\
$\lambda_\textrm{layers}$ & 6 & Defines the number of hidden layers.                  \\
$\lambda_\textrm{nodes}$ & 10 & Defines the number of hidden nodes per hidden layer.  \\
$\lambda_\textrm{gauss}$ & 1 & Defines the number of Gaussians used in the MDN. \\
$\lambda_\textrm{lr}$ & 0.0002 & Defines the learning rate of the Adam optimizer. \\
$\lambda_\textrm{wd}$ & 0.01 & Defines the weight-decay regularization. \\
$\lambda_\textrm{exp}$ & 506 & Defines the number of AME2020 data used in training. \\
$\lambda_\textrm{exp-pulls}$ & 50 & Defines the number of samples per AME mass. \\
$\lambda_\textrm{theory}$ & 50 & Defines the unique nuclei probed using the three models. \\
$\lambda_\textrm{theory-pulls}$ & 20 & Defines the number of samples per theory mass. \\
$\lambda_\textrm{physics}$  & 1.0 & Defines the strength of the physics loss enforcement. \\ 
$\lambda_\textrm{Z-low}$ & 5 & Defines the minimal proton number for the network. \\
$\lambda_\textrm{N-shells}$ & 8, 20, 28, 50, 82, 126, 184 & Defines the major closed neutron shells. \\
$\lambda_\textrm{Z-shells}$ & 8, 20, 28, 50, 82, 114 & Defines the major closed proton shells. \\
$\lambda_\textrm{FRDM2012}$ & 0.6 MeV & Defines the uncertainty used in probing FRDM2012 masses. \\
$\lambda_\textrm{WS4}$ & 0.5 MeV & Defines the uncertainty used in probing WS4 masses. \\
$\lambda_\textrm{HFB32}$ & 0.6 MeV & Defines the uncertainty used in probing HFB32 masses. \\

\end{tabular}
\end{table}

One essential observation of ground-state masses is that they obey the eponymous Garvey-Kelson (GK) relations \citep{Garvey1969}. 
This result suggests a judicious choice of mass differences of neighboring nuclei that minimizes the interactions between nucleons to first order, resulting in particular linear combinations that strategically sum to zero. 

If $N \geq Z$, the GK relations state that the mass difference is
\begin{equation}
\label{eqn:gk1}
\begin{aligned}
 & M_{Z-2,  N+2} - M_{Z, N} + M_{Z-1, N} \\
 & - M_{Z-2, N+1} + M_{Z, N+1} - M_{Z-1, N+2} \approx 0 \ ,
\end{aligned}
\end{equation}
and for $N < Z$, 
\begin{equation}
\label{eqn:gk2}
\begin{aligned}
 & M_{Z+2, N-2} - M_{Z, N} + M_{Z, N-1} \\
 & - M_{Z+1, N-2} + M_{Z+1, N} - M_{Z+2, N-1} \approx 0  \ .
\end{aligned}
\end{equation}
Higher order GK mass relations may also be considered, as in Ref.~\citet{Barea2008}. 
However, the use of these constraints alone does not yield viable predictions far from stability due to the accumulation of uncertainty as the relationship is recursively applied beyond known data \citep{Janecke1988}. 

As an alternative, we perform no such iteration in our application of the GK relations. 
Equations \ref{eqn:gk1} and \ref{eqn:gk2} are used directly, and it is important to recognize that  these equations depend exclusively on the masses.
Thus the second (physics-based) loss function can be defined purely as a function of the ML model's mass predictions. 

To enforce this physics-based observation, the second loss function can be defined as

\begin{equation}
    \mathcal{L}_2 = -\ln {\left ( \left| \sum_{\{C\}} GK(\bm{\mu}) \right| \right )} \ , 
    \label{eqn:ll_physics}
\end{equation}
where GK is function that defines the left-hand side of Equations \ref{eqn:gk1} and \ref{eqn:gk2} and we only use the model's predicted mean value of the masses, $\bm{\mu}$. 
The sum is performed over any choice of subset, $\{C\}$, of masses and does not have to overlap with the hybrid training data. 
The absolute value is necessary to ensure that the log-loss remains a real number. 
As with the data loss, $\mathcal{L}_1$, we also seek to simultaneously minimize the physics-loss, $\mathcal{L}_2$, which amounts to reducing the error among the difference in masses defined in the above equations.

An alternative to Equation \ref{eqn:ll_physics} that is potentially more restrictive, is to take the absolute value inside the summation
\begin{equation}
    \mathcal{L}^\textrm{alt}_2 = -\ln {\left ( \sum_{\{C\}} \left| GK(\bm{\mu} ) \right| \right )} \ .
    \label{eqn:ll_physics2}
\end{equation}
Because Equation \ref{eqn:ll_physics2} sums many non-zero items, it is a larger loss than using Equation \ref{eqn:ll_physics}. 
In this case, the strength of the physics hyperparameter (discussed below) should generally be lower than in the case of using Equation \ref{eqn:ll_physics}. 
A strong preference for selecting one functional form for the physics-based loss over the other has not been found. 

The total loss function used in training is taken as a sum of the data and physics losses
\begin{equation}
    \mathcal{L}_\textrm{total} = \mathcal{L}_1 + \lambda_\textrm{physics} \mathcal{L}_2
    \label{eqn:loss_total}
\end{equation}
where $\lambda_\textrm{physics}$ is a model hyperparameter which defines the strength of enforcement of the physics loss. 
We have found that values between 0.1 and 2 generally enforce the physics constraint in model predictions. 

\subsection{Assembling a model}

\begin{figure}[h!]
\begin{center}
\includegraphics[width=15cm]{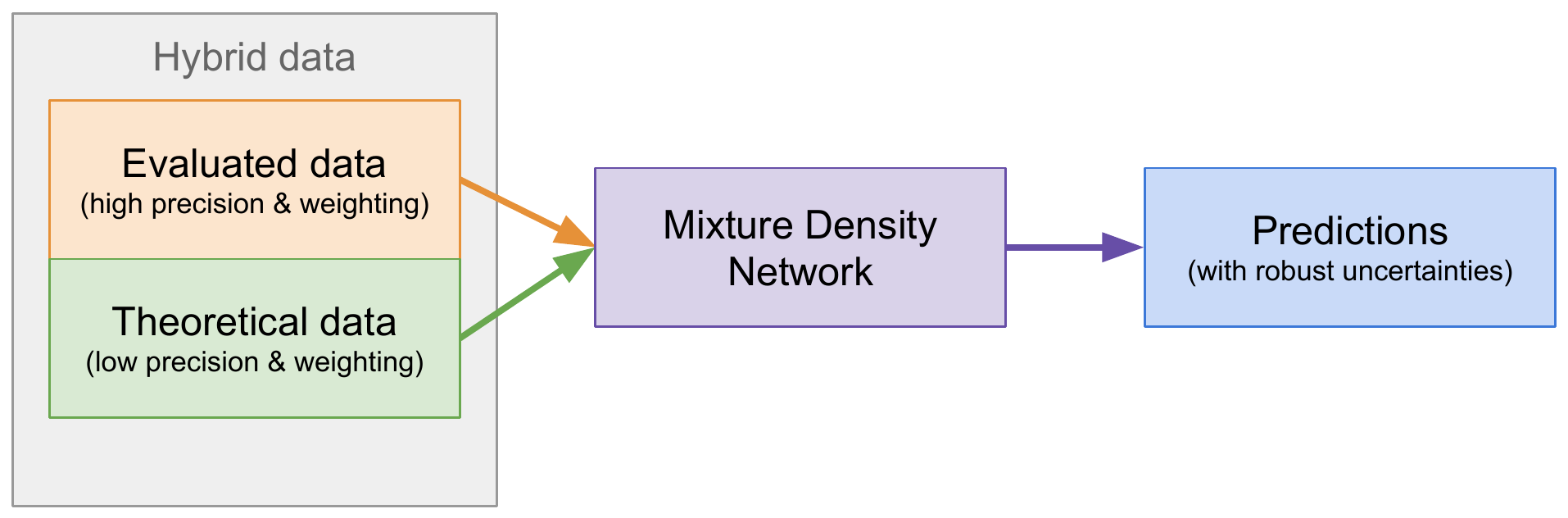}
\end{center}
\caption{A schematic of our methodology. The procedure used in this work combines high precision evaluated data with a handful of less-precise theoretical data. This results in predictions with well-quantified uncertainties across the chart of nuclides. }\label{fig:schematic}
\end{figure}

\begin{figure}[h!]
\begin{center}
\includegraphics[width=15cm]{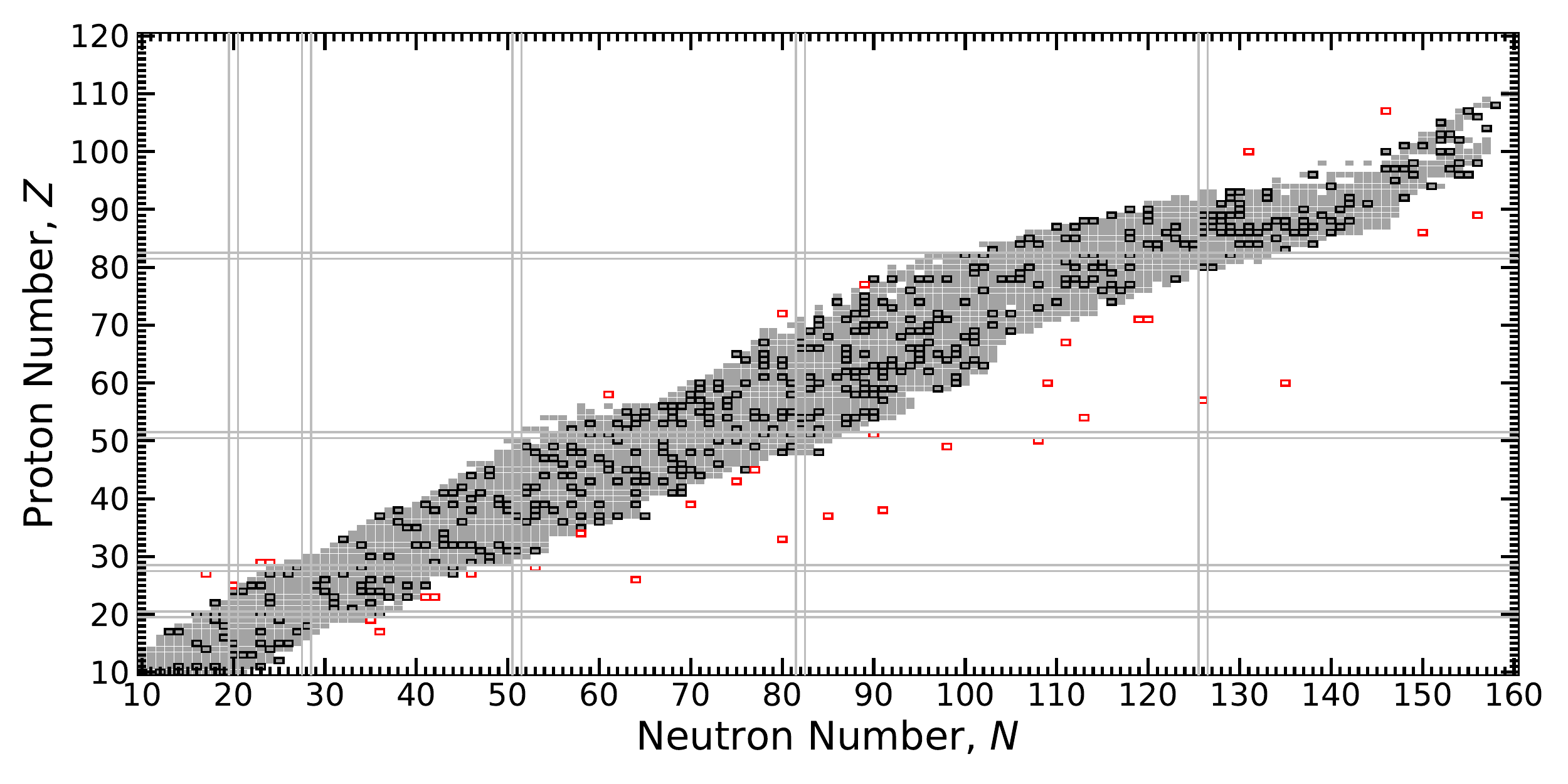}
\end{center}
\caption{The chart of nuclides showing the extent of the 2020 Atomic Mass Evaluation (AME) indicated by grey squares, training nuclei part of the AME indicated by black squares, and the extra theoretical nuclei indicated by red squares. Closed proton and neutron shells are indicated by parallel lines. }\label{fig:training}
\end{figure}

A schematic of our methodology is shown in Figure~\ref{fig:schematic} and encapsulated below. 
The modeling of masses begins by setting hyperparameters, summarized in Table~\ref{tab:hyperparams}, for the particular calculation. 
A random selection of hybrid data is made, as can be seen in Figure~\ref{fig:training}. 
The bulk of the masses selected for training comes from the AME (black squares) where high-precision evaluated data resides (gray squares). 
Only a handful of masses from theoretical models are taken for training (red squares). 

After selection of hyperparameters and data, training begins which seeks to minimize the total loss, Eqn.~\ref{eqn:loss_total}. 
Training can take many epochs, and the data loss as well as the physics loss play important roles throughout this process, as discussed in \citet{Mumpower2022}. 
Once the MDN has been trained on data, the results are assembled into predictions by sampling the posterior distribution several thousand times. 
The final output is a prediction of the mean value of the expected mass, $M$, and its associated uncertainty, $\sigma(M)$ for any provided nucleus defined by ($Z$, $N$). 

\section{Results}
\label{sec:results}

In this section we present a MDN model trained on hybrid data. 
We analyze the performance with known data and discern the ability to extrapolate model predictions. 
We evaluate the impact of including theoretical data and the physics-based loss function. 

\subsection{Comparison with existing data}

The final match to 506 training nuclei for our MDN model is $\sigma_\textrm{RMS} = 0.279$ MeV. 
The total $\sigma_\textrm{RMS}$ for the entire AME2020 is $0.395$ MeV. 
This is an increase of roughly $0.116$ MeV between training and verification data which is on the order of the accuracy of the GK relations. 
We limit these calculations to nuclei with $A \geq 50$. 
While the model can predict masses for nuclei lighter than $A=50$, it generally performs worse in this region because there are fewer nuclei at lower mass numbers than in heavier mass regions. 
Therefore, there are fewer light nuclei selected in the random sample than heavy nuclei. 

\begin{figure}[h!]
\begin{center}
\includegraphics[width=15cm]{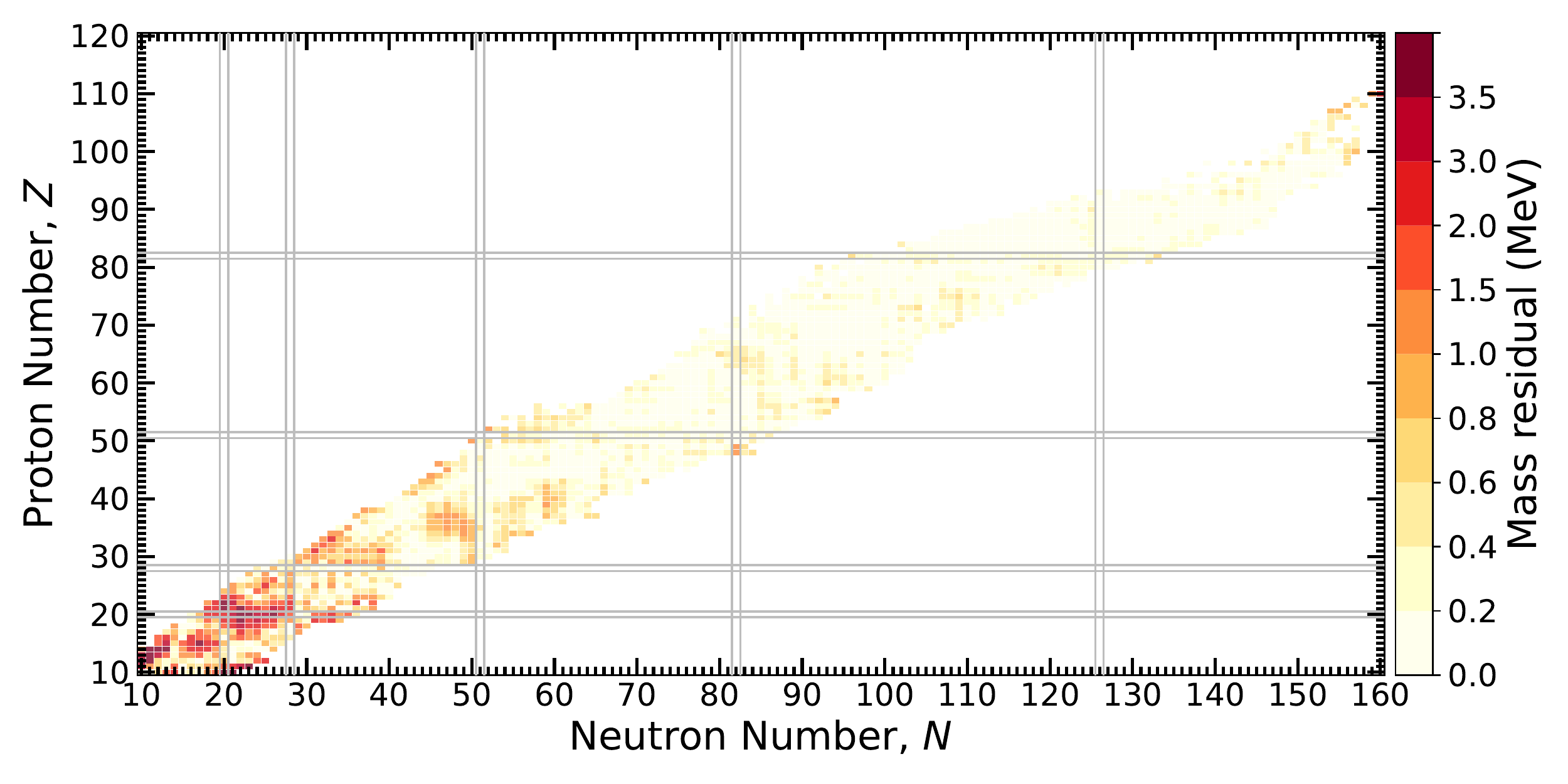}
\end{center}
\caption{The absolute value of mass residuals across the chart of nuclides using an MDN model and the AME2020. Heavier nuclei are generally well described by the model while lighter nuclei exhibit larger discrepancies. See text for details. }\label{fig:residuals}
\end{figure}

The absolute value of the mass residual, $\Delta M_{Z,N} = | M^\textrm{AME}_{Z,N} - M^\textrm{MDN}_{Z,N} |$, is one way to measure model performance. 
Figure~\ref{fig:residuals} plots this quantity across the chart of nuclides versus the AME2020. 
The predictions of light nuclei tend to have an error on the order of several MeV with heavier nuclei around $0.3$ MeV. 
The MDN model performance is on par with commonly used models in the literature. 

In comparison to our previous results discussed in \citet{Mumpower2022}, the addition of light nuclei in training is found to relatively increase the discrepancy for heavier nuclei. 
The additional information modestly reduces the overall model quality as measured by $\sigma_\textrm{RMS}$. 
On the other hand, the model is better positioned to describe the nuclear landscape more completely, insofar as the training process introduces information on the nuclear interaction that is uniquely captured in low-mass systems.

\begin{figure}[h!]
\begin{center}
\includegraphics[width=15cm]{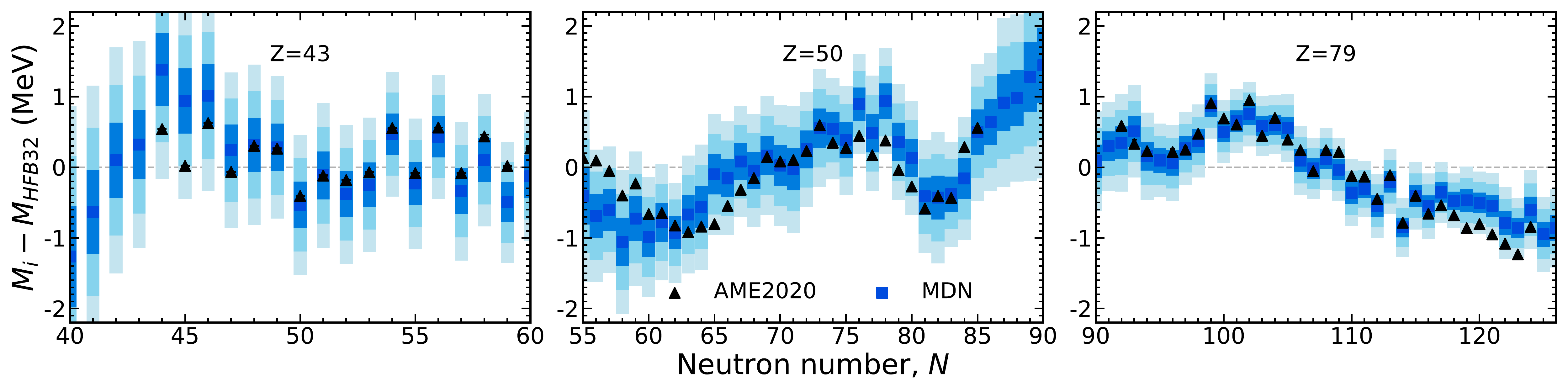}
\end{center}
\caption{The prediction of masses along three isotopic chains in comparison to AME2020 data. Masses are plotted in reference to HFB32. The MDN model captures the trends exhibited in data and furnishes individual uncertainties (the one, two, and, three sigma confidence intervals are shown by blue shading). }\label{fig:masses_isotopic}
\end{figure}

The behavior of the model with respect to select isotopic chains are shown in Figure~\ref{fig:masses_isotopic}. 
In regions of the chart where the MDN model is confident in its predictions such as in the $Z=79$ isotopic chain, the uncertainties are very well constrained. 
The converse is also true, as is the case with higher uncertainties along the $Z=43$ isotopic chain. 
The tin ($Z=50$) isotopic chain highlights an intermediate case. 

Inspection of this figure shows that the model has a preference for evaluated data in this region and does not follow the trends of HFB32, despite HFB32 masses being provided for training. 
This result reveals a unique feature of our modeling: evaluated data, due to its low uncertainties, is highly favored while theoretical points, with relatively larger uncertainties, are used as guideposts for how nuclei behave where there is no data. 
How much a particularly model is favored farther from stability depends on how much weighting we provide it with the choice of model uncertainty. 
The trends of the MDN predictions are discussed in the next section. 

\subsection{Trends away from measured data}

The extrapolation quality of atomic mass predictions is an important problem, especially for astrophysical applications where this information is needed for thousands of unmeasured species \citep{Mumpower2016, Martin2016}. 
The formation of the elements in particular requires robust predictions with well-quantified uncertainties \citep{Sprouse2020}. 
The MDN supplies such information, and we now gauge the quality of the extrapolations by comparison with other theoretical models. 

\begin{figure}[h!]
\begin{center}
\includegraphics[width=15cm]{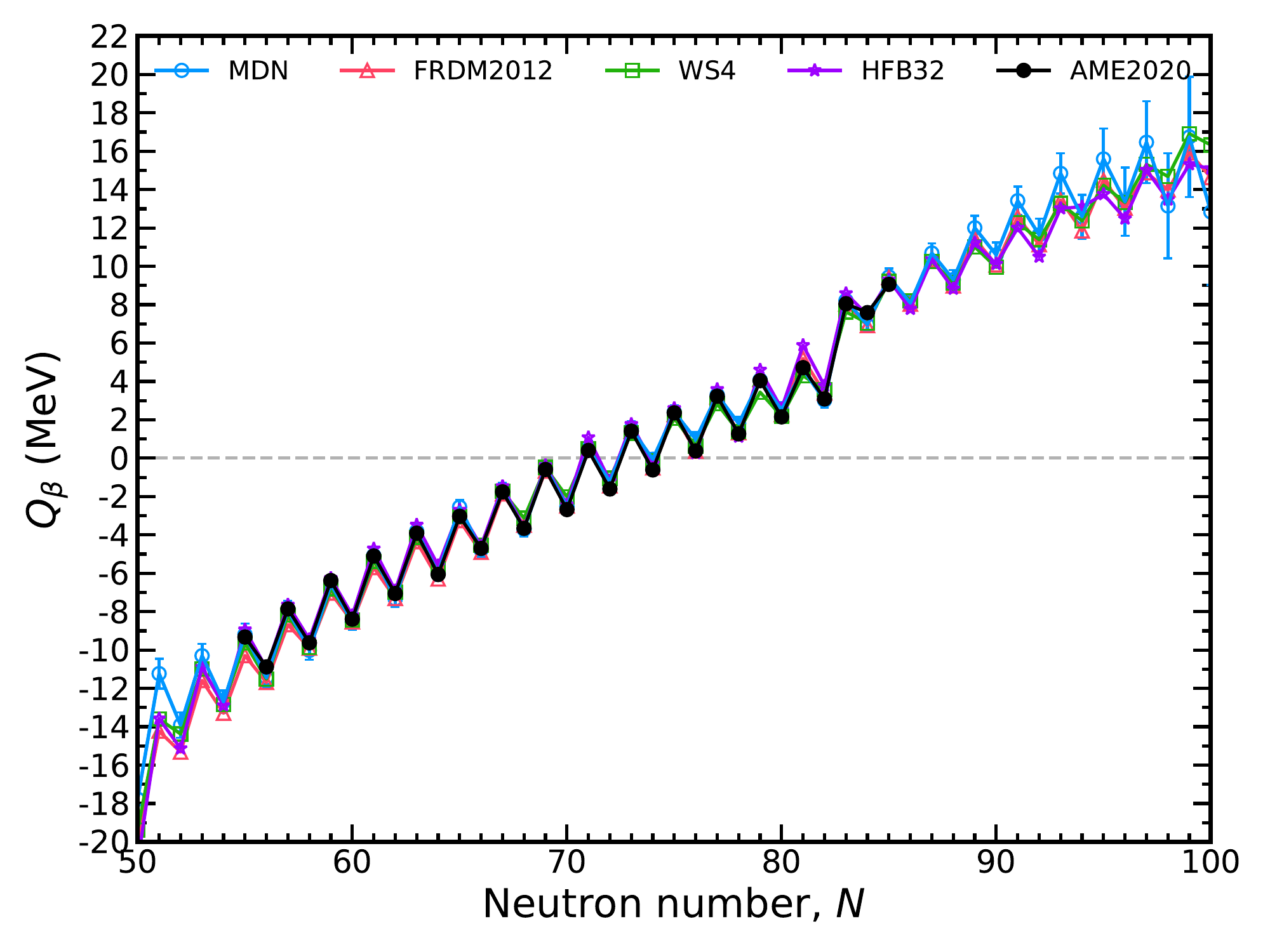}
\end{center}
\caption{The total available energy for nuclear $\beta^{-}$ decay, $Q_\beta$, along the tin isotopic chain ($Z=50$) with a 1-$\sigma$ confidence interval. The MDN model (blue) reproduces known data (black) and continues reasonably physical behavior when extrapolating. Theoretical models used in training are plotted for comparison. }\label{fig:qbm}
\end{figure}

The $Q_\beta(Z,N) = M_{Z,N} - M_{Z+1, N-1}$ is plotted in Figure~\ref{fig:qbm} for the tin ($Z=50$) isotopic chain. 
Additionally plotted is the uncertainty, $\delta Q_\beta$, which is calculated via propagation of error
\begin{equation}
    \delta Q_\beta(Z,N) = \sqrt{ \sigma^2(M_{Z,N}) + \sigma^2(M_{Z+1, N-1}) - 2\sigma(M_{Z,N}, M_{Z+1, N-1}) } \ . 
    \label{eqn:qbm_uncert}
\end{equation}
The mass uncertainties $\sigma(M_{Z,N})$, $\sigma(M_{Z+1, N-1})$ are outputs of the MDN. 
The correlation between the masses, $\sigma(M_{Z,N}, M_{Z+1, N-1})$, is assumed to be zero. 
The model has excellent performance where data is known and this result can be considered representative for other isotopic chains. 
Predicted uncertainties grow with decreasing and increasing neutron number outside of measured data, underscoring the Bayesian nature of our approach. 

Also shown in Figure~\ref{fig:qbm} are the theoretical models used in training. 
Comparison with these models shows that the MDN continues to retain physical behavior when extrapolating to neutron deficient or neutron rich regions. 
Intriguingly, the MDN model does not preferentiate one specific model when extrapolating. 
Instead, where there begins to be discrepancies among the theoretical models, the uncertainties begin to increase. 
For $Q_\beta$, the predictions along the tin isotopic chain begin to be dominated by uncertainties roughly ten units from the last available AME2020 data point. 

\begin{figure}[h!]
\begin{center}
\includegraphics[width=15cm]{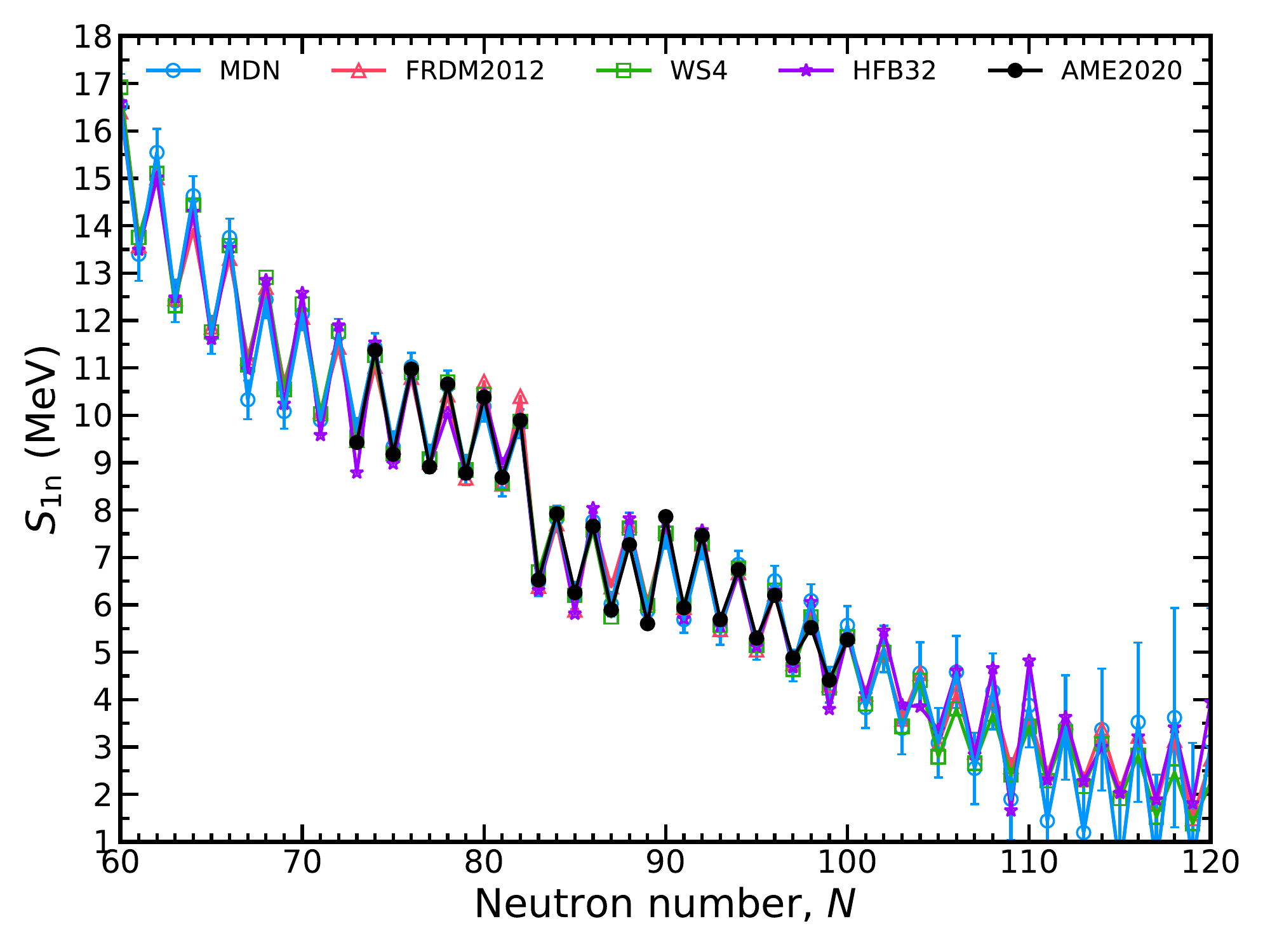}
\end{center}
\caption{The one-neutron separation energy, $S_{1\textrm{n}}$, along the promethium isotopic chain ($Z=61$) with a 1-$\sigma$ confidence interval. The MDN model (blue) reproduces known data (black) and continues reasonably physical behavior when extrapolating. Theoretical models used in training are plotted for comparison. }\label{fig:s1n}
\end{figure}

In Figure~\ref{fig:s1n} we show the extrapolation quality of one-neutron separation energies, $S_{1\textrm{n}}(Z,N) = M_n + M_{Z,N-1} - M_{Z,N}$. 
The propagation of error formula, Eqn.~\ref{eqn:qbm_uncert}, is again employed to calculate $\delta S_{1\textrm{n}}$ since this quantity also depends on mass differences. 
The uncertainty of the mass of the neutron, $M_n$, can be safely ignored. 
The qualitative behavior of $S_{1\textrm{n}}$ is well described. 
No unphysical inversions of $S_{1\textrm{n}}$ are found with the MDN model in contrast to the HFB32 model where this behavior can arise; observe around $N=105$. 
Again, we find that roughly ten units from the last measured isotope, uncertainties begin to rise substantially. 

A consequence of the growth of uncertainties is that the prediction of the neutron dripline, $S_{1\textrm{n}}=0$ is also largely uncertain for any isotopic chain. 
We conclude that hybrid data does not presently place stringent constraints on this quantity, which is widely recognized as an open problem by the community \citep{Erler2012, Xia2018, Neufcourt2020}. 

\begin{figure}[h!]
\begin{center}
\includegraphics[width=15cm]{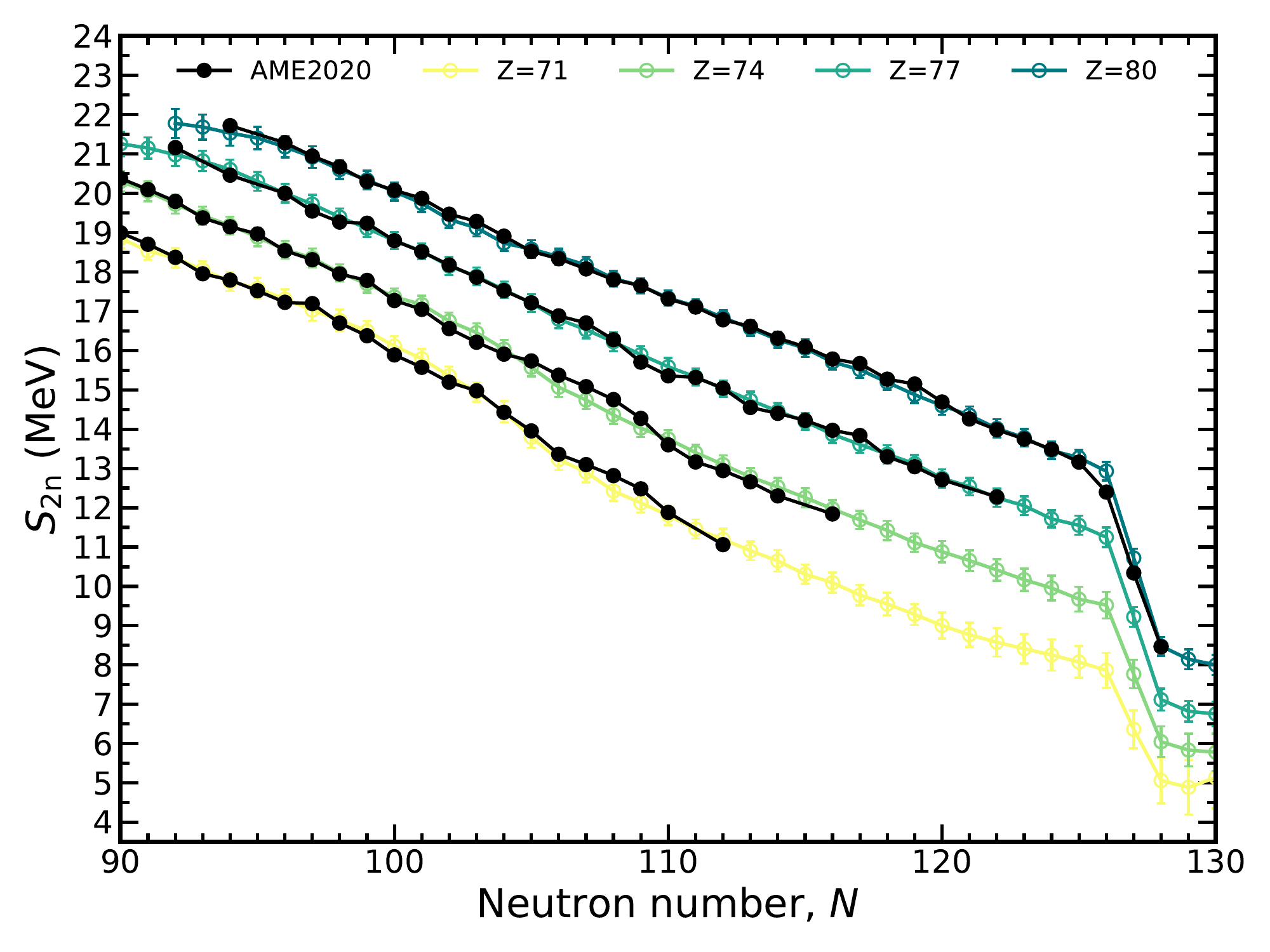}
\end{center}
\caption{The two-neutron separation energy, $S_{2\textrm{n}}$, along several isotopic chains plotted with a 1-$\sigma$ confidence interval. The MDN model (colors) reproduces known data (black) and continues reasonably physical behavior when extrapolating. }\label{fig:s2n}
\end{figure}

Another quantity that can be used to gauge the quality of extrapolations is the two-neutron separation energy, $S_{2\textrm{n}}(Z,N) = 2M_n + M_{Z,N-2} - M_{Z,N}$. 
The two-neutron separation exhibits less odd-even staggering than $S_{1\textrm{n}}$ because the subtraction always pairs even-N or odd-N nuclei. 
The behavior of the MDN model is shown in Figure~\ref{fig:s2n} for the lutetium ($Z=71$), tungsten ($Z=74$), iridium ($Z=71$), and mercury ($Z=80$) isotopic chains. 
All experimental data falls within the 1-$\sigma$ confidence intervals except for $^{206}$Hg. 
A relatively robust shell closure is predicted at $N=126$, though there is some weakening at the smaller proton numbers. 

\begin{figure}[h!]
\begin{center}
\includegraphics[width=15cm]{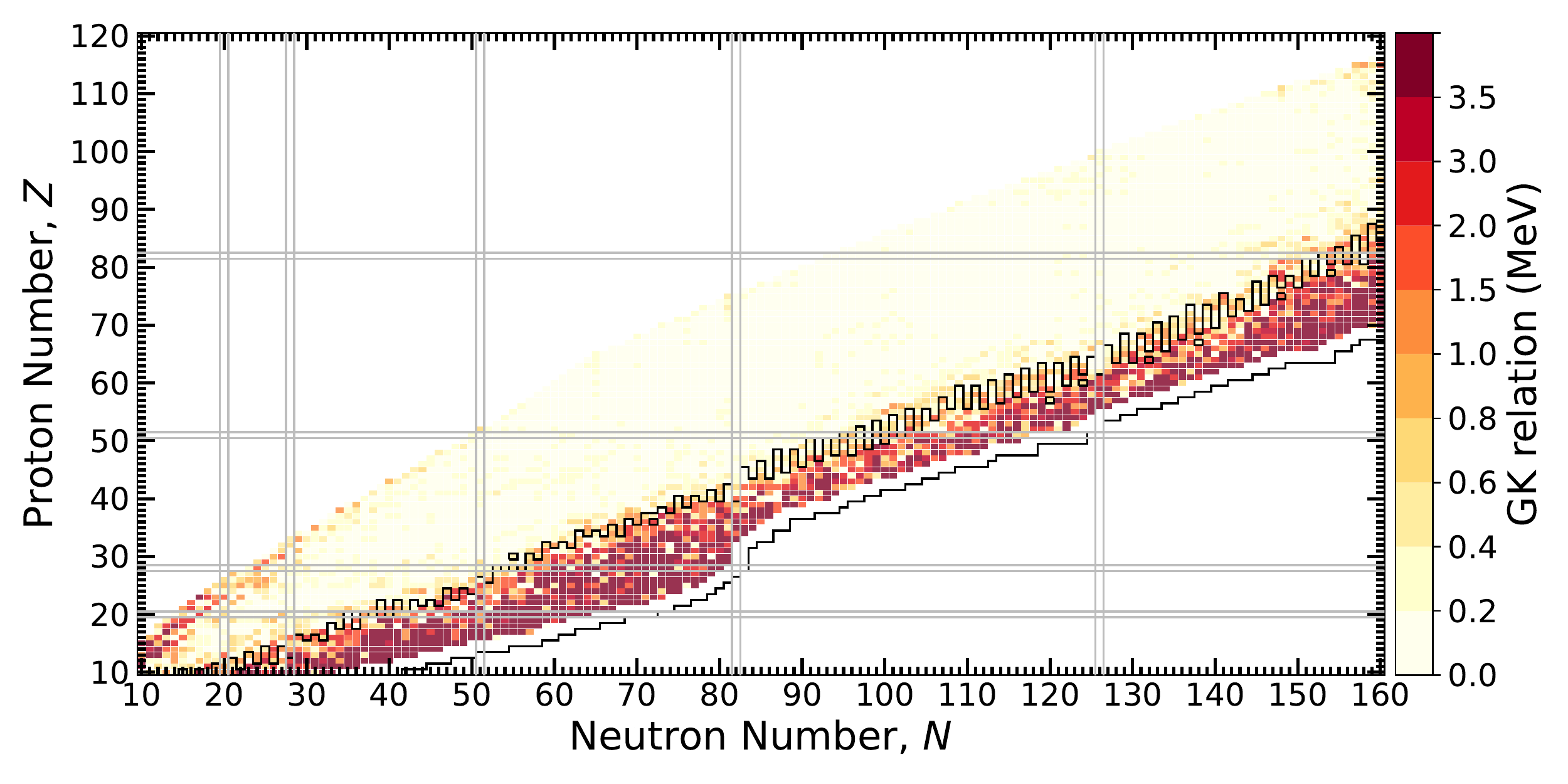}
\end{center}
\caption{A test of how well the GK relations are maintained throughout the chart of nuclides. Lower values indicate predictions inline with GK, which is found nearly everywhere, except for the most extreme cases where the model is uncertain at the limits of bound nuclei. The black outlined nuclei have $\delta M > S_{1\textrm{n}}$, indicating where mass uncertainties are large.   }\label{fig:gk_test}
\end{figure}

Finally, we consider the behavior of the physics of ground-state masses across the nuclear chart using the predictions of the MDN model. 
Whether or not the GK relations are preserved is yet another test of the extrapolation quality of the MDN. 
The calculation of the left hand side of Eqns.~\ref{eqn:gk1} and \ref{eqn:gk2} is shown in Figure~\ref{fig:gk_test} across the chart of nuclides. 
Yellow shading indicates the GK relations are satisfied while orange and red shading indicate potential problem areas where the relations have broken down. 
Given the behavior seen in the previous figures, it is clear that towards the neutron dripline, the uncertainties have grown so large that the model is unsure of the preservation of the GK relations. 
To emphasize this point, we calculate the nuclei for which the mass uncertainty, $\delta M$, is larger than the one-neutron separation energy, $S_{1\textrm{n}}$, and designate this as a region bound in black. 
We observe that this bounded region is precisely where the orange and red regions are located. 
Figure~\ref{fig:gk_test} suggests that a potential modification of the loss function that encodes the GK relations could be made to include uncertainties obtained from the MDN. 
This line of reasoning is the subject of future work. 

\subsection{Impact of theoretical data and physics constraint}

\begin{figure}[h!]
\begin{center}
\includegraphics[width=15cm]{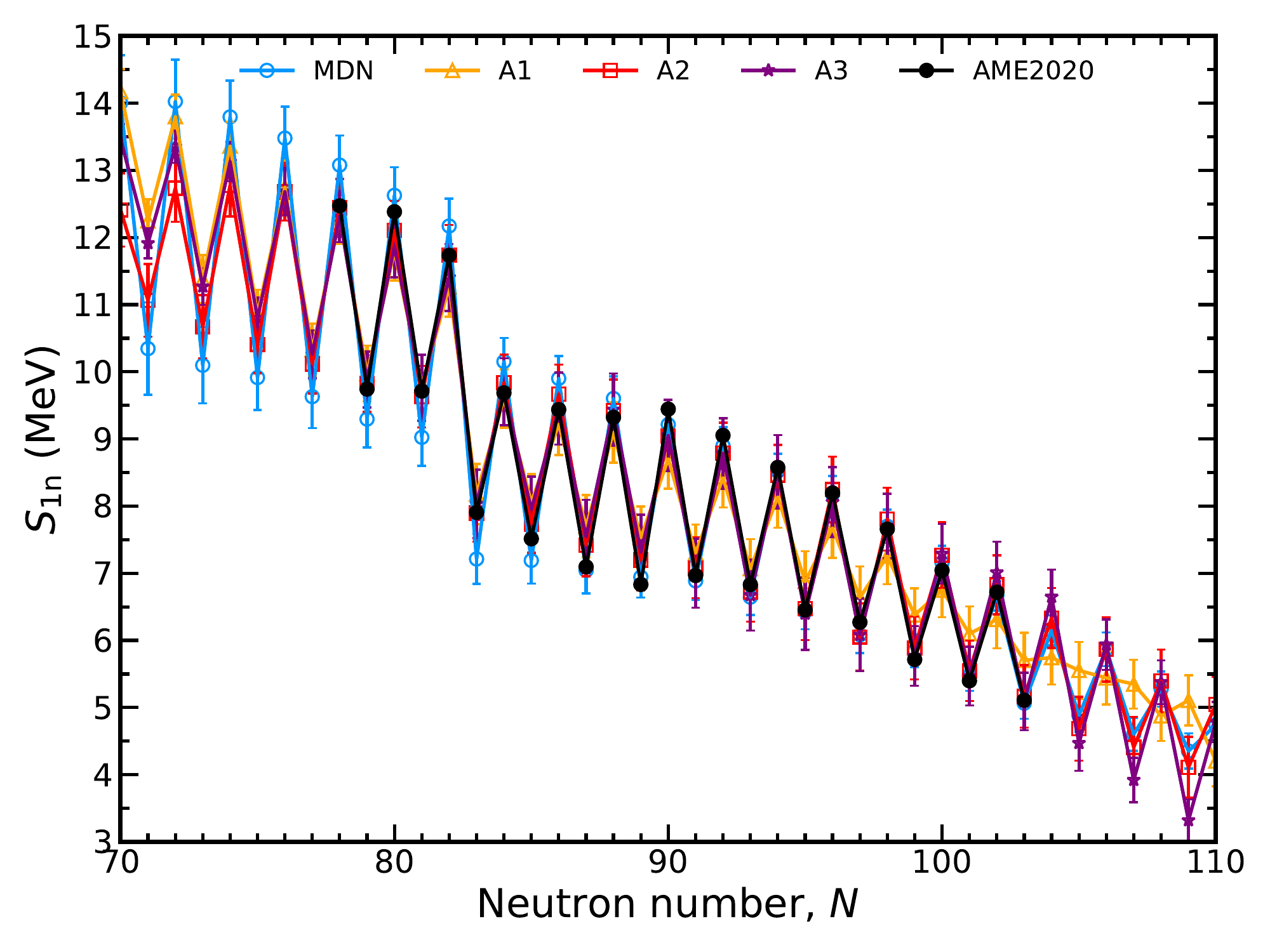}
\end{center}
\caption{Comparison of MDN models with different assumptions for input data and training loss along the dysprosium ($Z=66$) isotopic chain. See text for details. }\label{fig:model_comparison_s1n}
\end{figure}

We now assess the impact of the inclusion of theoretical data and the physics loss on the predictive capabilities of the MDN. 
Figure~\ref{fig:model_comparison_s1n} shows four different training sets in the context of $S_{1\textrm{n}}$ values. 
The line labeled MDN is the network shown throughout this manuscript that includes both hybrid data and physical constraint.  
A1 is a MDN model trained only with experimental data, lacking information about theory or the physical loss defined by the GK relations;
A2 is a MDN model trained with the physics loss but without theoretical data; and A3 is a MDN model trained with theory data but without any physics loss.

From these four sets, it is clear that both hybrid data \textit{and} the physics-based loss are necessary to provide quality extrapolations into unknown regions. 
Training with the lack of theory and GK (A1) exhibits a less desirable preference for a smooth extrapolation of $S_{1\textrm{n}}$ values. 
The addition of the physics loss (A2) improves the situation by restoring the odd-even behavior observed in measured nuclei. 
The expected behavior in $S_{1\textrm{n}}$ is also restored by run A3, where the hybrid data includes theory but training is not informed of the GK relations. 

We note that the improvement in extrapolation behavior resulting from the hybrid data and physics-based loss is generally independent of any hyperparameters that otherwise appear in the MDN. 
In particular, we have preliminarily verified this result against systematic variations in relevant hyperparameters, including network size, training size, different nuclei, different input features and different blends of theoretical models. 
All of these variations demonstrate a similar propensity for improvement when both physics loss \textit{and} hybrid data are used. 
These results lead us to reaffirm our previous observation that the addition of theory data serves as guideposts for the network solutions while the GK relations are used to ensure a refined solution. 

\subsection{Estimated growth of uncertainties with neutron number}

\begin{figure}[h!]
\begin{center}
\includegraphics[width=15cm]{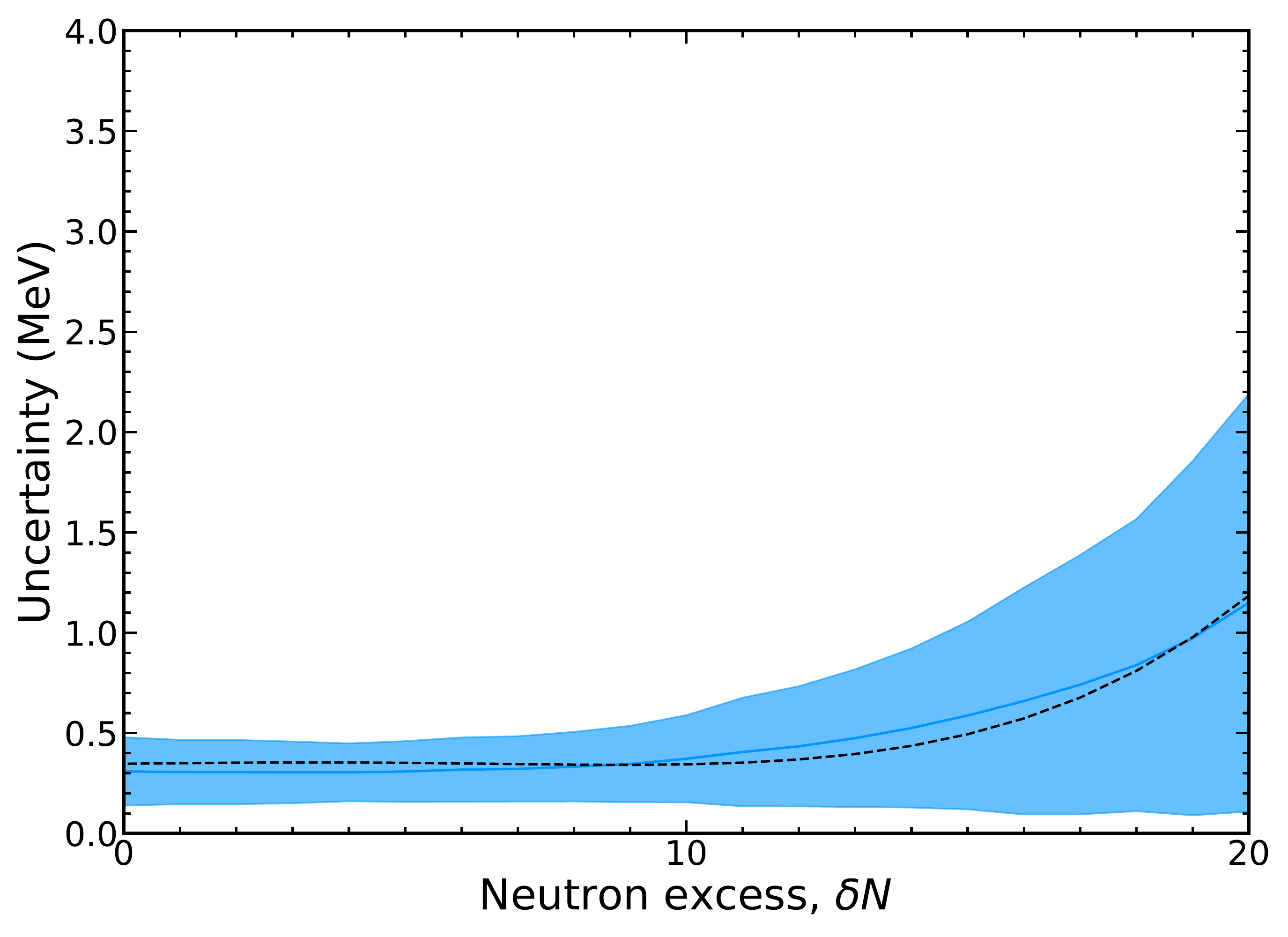}
\end{center}
\caption{The average and standard deviation of the growth of mass uncertainties as a function of neutron excess. The functional form of the average can be well modeled by Eqn.~\ref{eqn:uncert_nexcess} as plotted with a dashed black line. }\label{fig:nexcess_uncert}
\end{figure}

The behavior of mass uncertainties as one traverses the chart of nuclides can be ascertained using the MDN predictions. 
Here we consider the evolution of uncertainties provided by an MDN model as a function of neutron excess, $\delta N$. 
We take the definition of $\delta N$ to mean the number of neutrons from the last stable isotope and use the NuBase (2020) data to make this determination \citep{Kondev2021}. 
For each isotopic chain, $\delta N$ may reference a slightly different neutron number for the particular isotope. 
The choice of this variable provides a relatively straight forward way to observe how mass uncertainties grow far from known data.  
The average and standard deviation of the MDN model's uncertainties are shown in Fig.~\ref{fig:nexcess_uncert}. 

The functional form of the average uncertainty growth as a function of neutron excess is well approximated by the following relation, 
\begin{equation}
\sigma(\delta N) \approx p_0 + 
                         p_1\delta N + 
                         p_2\delta N^2 + 
                         p_3\delta N^3 + 
                         p_4\delta N^4 + 
                         p_5^{\delta N}
                         \ ,
\label{eqn:uncert_nexcess}
\end{equation}
where the parameters, $p_{i}$, are fit numerically (least squares) and are given in Table~\ref{tab:nexcess_uncert}. 
This functional form may be readily used in simulations of nucleosynthesis to approximate uncertainties in masses with models that do not provide this information. 

\begin{table}[]
\caption{The parameters found using a least-square fit of Eqn.~\ref{eqn:uncert_nexcess} to the MDN uncertainties across the chart of nuclides. }
\label{tab:nexcess_uncert}
\begin{tabular}{cccccc}
\textbf{$p_0$} & \textbf{$p_1$} & \textbf{$p_2$} & \textbf{$p_3$} & \textbf{$p_4$} & \textbf{$p_5$} \\
$-7.035 \times 10^{-1}$ & $-1.037 \times 10^{-1}$ & $-2.997 \times 10^{-3}$ & $-3.729 \times 10^{-4}$ & $8.782 \times 10^{-7}$ & $1.109 \times 10^{0}$ \\
\end{tabular}
\end{table}

\section{Conclusion}
\label{sec:conslusion}

We present a Bayesian averaging technique that can be used to study the ground-state masses of atomic nuclei with corresponding uncertainties. 
In this work, we combine high-precision evaluated data, weighted strongly, with theoretical data for nuclei which are further from stability, more poorly understood, and therefore weighted more weakly. 
Training of a probabilistic neural network is used to construct the posterior distribution of ground-state masses. 
Along with a loss function for matching data, a second, physics-based loss function is employed in training to emphasize the relevant local behavior of masses. 
Excellent performance is obtained with comparison to known data, on the order of $\sigma_\textrm{RMS} \sim 0.3$ MeV, and the physical behavior of solutions is maintained when extrapolating. 
It is found that available data from experiment and theory are not, at this time, sufficient to resolve the relatively large uncertainties towards the limits of bound nuclei using the framework developed in this work. We emphasize the continuing need for advances in nuclear experiment and theory to reduce these uncertainties.

Our Bayesian averaging procedure enables the rapid construction of a mass model using any combination of precise and imprecise data through adjustable stochastic weights of the hybrid training inputs. 
For instance, if a particular theoretical model is favored over another, its sampling can be adjusted accordingly to emphasize its importance. 
Similarly, new high-precision data may be incorporated in the future from measurement campaigns at radioactive beam facilities. 
At the same time, our technique also enables freedom in the exploration of the relevant physics of ground-state masses. 
This can be achieved by probing a variety of physics-based features, for example, or by introducing alternative physics-based loss functions in training. 

The methodology outlined here can be generalized to describe any nuclear physics property of interest, particularly when reliable extrapolations are necessary. 
This technique opens new avenues into Machine Learning research in the context of nuclear physics through the unification of data, theory, and associated physical constraints to empower predictions with quantified uncertainties. 
We look forward to extensions of this work to model nuclear decay properties, such as half-lives and branching ratios, as particularly promising opportunities in the near future.

\section*{Conflict of Interest Statement}

The authors declare that the research was conducted in the absence of any commercial or financial relationships that could be construed as a potential conflict of interest.

\section*{Author Contributions}

M.~R.~Mumpower led the study, overseeing all aspects of the project, including writing the code used in this study. 
M.~Li ran and analyzed combinations of hybrid input data as well as hyperparamters. 
T.~M.~Sprouse, A.~E.~Lovell, and A.~T.~Mohan provided technical expertise for Machine Learning methods. 
All authors contributed to the writing of the manuscript, and read and approved the submitted version.

\section*{Funding}
This research was supported by the Los Alamos National Laboratory (LANL) through its Center for Space and Earth Science (CSES). 
CSES is funded by LANL’s Laboratory Directed Research and Development (LDRD) program under project number 20210528CR. 

\section*{Acknowledgments}
M.R.M., M.L., T.M.S., A.E.L., A.T.M. were supported by the US Department of Energy through the Los Alamos National Laboratory (LANL). LANL is operated by Triad National Security, LLC, for the National Nuclear Security Administration of U.S.\ Department of Energy (Contract No.\ 89233218CNA000001). 
M. L. and B. S. M. were supported by NASA Emerging Worlds grant 80NSSC20K0338. 


\section*{Data Availability Statement}
The raw data supporting the conclusions of this article will be made available by the authors, without undue reservation. 

\bibliography{refs}

\end{document}